\documentclass[%
 aip,
 amsmath,amssymb,
 reprint,%
]{revtex4-1}

\usepackage[dvipdfmx]{graphicx}

\usepackage[utf8]{inputenc}
\usepackage[T1]{fontenc}
\usepackage{mathptmx}
\usepackage{etoolbox}
\usepackage{braket}
\usepackage{color}
\usepackage{ulem}

\makeatletter
\def\@email#1#2{%
 \endgroup
 \patchcmd{\titleblock@produce}
  {\frontmatter@RRAPformat}
  {\frontmatter@RRAPformat{\produce@RRAP{*#1\href{mailto:#2}{#2}}}\frontmatter@RRAPformat}
  {}{}
}%
\makeatother
\begin{document}

\preprint{AIP/123-QED}

\title[Sample title]{{\it{Ab initio}} study on the possible magnetic topological semimetallic state \\ in MnMg$_{2}$O$_{4}$}
\author{Satoshi Tomita}
\affiliation{ 
Department of Applied Physics, Tohoku University, Sendai 980-8579, Japan 
}%
\author{DaPeng Yao}%
\affiliation{ 
Department of Physics, Tokyo Institute of Technology, Tokyo 152-8551, Japan 
}%
\author{Hiroki Tsuchiura}%
 \email{hiroki.tsuchiura.e8@tohoku.ac.jp .}
 \altaffiliation[Also at ]{Center for Spintronics Research Network, Tohoku University, Sendai 980-8577, Japan.}
\affiliation{ 
Department of Applied Physics, Tohoku University, Sendai 980-8579, Japan 
}%

\author{Kentaro Nomura}
 \altaffiliation[Also at ]{Center for Spintronics Research Network, Tohoku University, Sendai 980-8577, Japan.}
\affiliation{%
Institute for Materials Research, Tohoku University, Sendai 980-8577, Japan 
}%

\date{\today}

\begin{abstract}
We study the electronic state of an inverse spinel compound MnMg$_{2}$O$_{4}$ based on
first-principles calculations.
The high-spin state is realized in Mn ions on the diamond lattice, resulting in that this material is found to be a half-metallic semimetal with the minority spin-gap about 3eV,
and also with line nodes in the Brillouin zone.
The intrinsic anomalous Hall conductivity (AHC) is also computed as a function of the chemical potential of the system assuming the rigid band structure, and is found to exhibit a peak structure with a maximum value of 200 S/cm at only 15 meV above the Fermi level.
The relation between the large AHC and Berry curvature in the Brillouin zone is also discussed.

\end{abstract}

\maketitle


\section{INTRODUCTION}

Topological semimetals, such as Weyl semimetals, are in a class of gapless electronic
phases that exhibit topologically stable crossings of energy bands. 
In these materials, Berry curvature emerging within the band structure near 
the Fermi energy acts like a magnetic field in momentum space, resulting in robust
bulk transverse transport against disorder.
It provides novel degrees of freedom in 
spintronic applications or thermoelectric
energy conversion.
Thus the search for topological semimetals has been one of the most intriguing issues
in the field of spintronics.

Theoretically, a spinel compound HgCr$_{2}$Se$_{4}$ was the first to be predicted 
to be a magnetic Weyl semimetal\cite{HgCr_2Se_4}.
Several years later, soon after the experimental confirmation of a theoretical prediction 
that TaAs is a nonmagnetic Weyl semimetal\cite{TaAs_theory}, a few magnetic
materials such as 
a shandite compound Co$_{3}$Sn$_{2}$S$_{2}$ and the (inverse) Heusler compounds
Co$_{2}$MnAl and Ti$_{2}$MnAl were theoretically predicted to be magnetic Weyl semimetals, 
some of which were confirmed experimentally\cite{TaAs,Co3Sn2S2_1,Co3Sn2S2_2,Co3Sn2S2_3,Co2MnAl,Ti2MnAl}
.

Quite recently, the inverse spinel compound VMg$_{2}$O$_{4}$ has been predicted to be an excellent
topological semimetal with the so-called hourglass-shaped band dispersion based on
an effective tight-binding model and also on first-principles calculations
\cite{Jiang,Zhang}. 
The hourglass-shaped band dispersion is formed on the $e_{g}$ band consisting of V 3d orbitals
on the diamond lattice, which is fully polarized and well separated from the other bands.
It is also found that VMg$_{2}$O$_{4}$ becomes a magnetic Weyl semimetal when the spin-orbit interaction is taken into account. 
Since spinel compounds have been widely studied from various aspects in the field of 
spintronics, for example, the application of a nonmagnetic spinel MgAl$_{2}$O$_{4}$ as a tunneling barrier of tunnel magneto-resistance device
\cite{spinel_barrier_1,spinel_barrier_2}
, and the utilization of strain-induced
high magnetic anisotropy in (CoFe)$_{2}$O$_{4}$ ferrite
\cite{co_ferrite},
the search for new topological semimetals by shedding the light of topological physics
on spinel compounds may open the door to the invention of novel spintonic devices.

Let us now look at the 3d electronic states of VMg$_{2}$O$_{4}$.
If each V ion maintains the nominal V$^{4+}$ charge, then it has one electron on the 
$e_{g}$-orbitals. 
In a tight-binding picture, this corresponds to the situation where there is one electron
per each site of the diamond lattice which consists of two degenerate orbitals.
Here we may ask what happens if we consider a similar situation with $t_{2g}$-orbitals.
Such a situation can be realized in MnMg$_{2}$O$_{4}$ in which V is totally substituted
for Mn.
If the high-spin state is realized in the tetravalent configuration of Mn ions, 
there is one electron in the $t_{2g}$-orbitals.
Thus, 
we can expect that this system 
is a $t_{2g}$ counterpart of VMg$_{2}$O$_{4}$.

In this paper, we study the electronic state of MnMg$_{2}$O$_{4}$, and examine the topological nature of the transverse transport properties of this system by calculating the intrinsic anomalous Hall conductivity (AHC) and the electronic Berry curvature over the occupied portions of the Brillouin zone based on first-principles calculations.

\section{Computational details}

\begin{figure}
\includegraphics[clip,width=7cm]{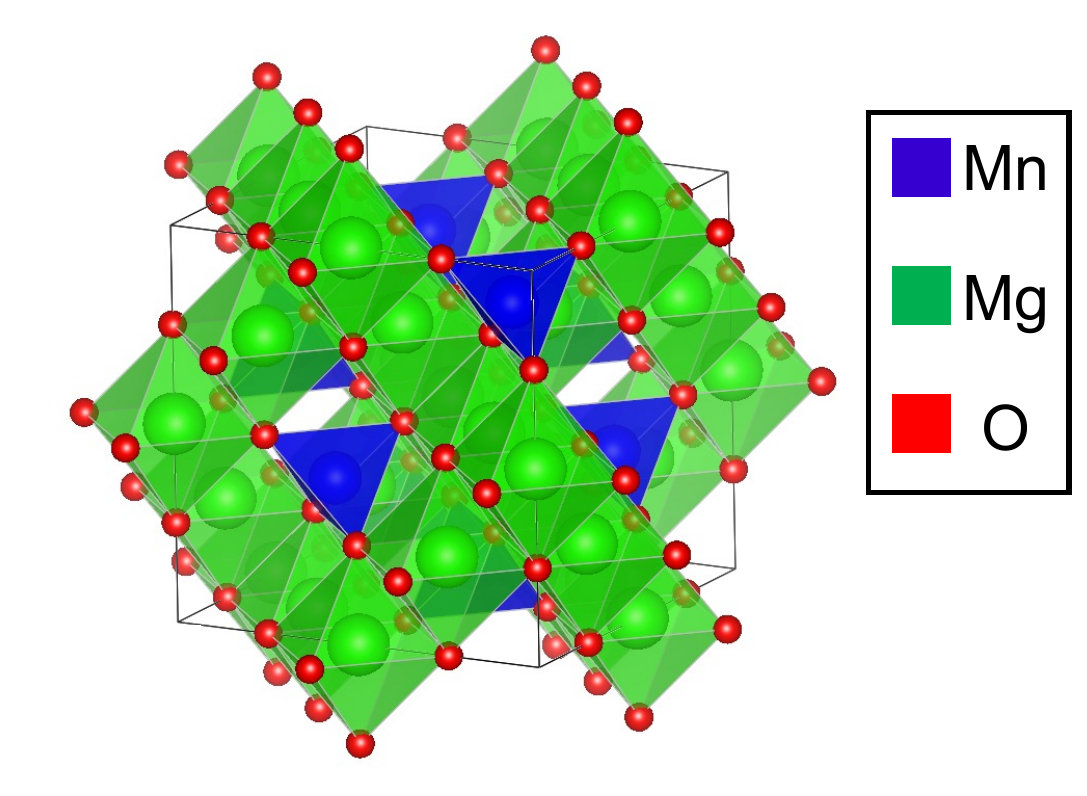}
\caption{\label{fig:epsart} Conventinal unit cell for MnMg$_{2}$O$_{4}$.}
\label{struct}
\end{figure}
To investigate the electronic band structure of MnMg$_{2}$O$_{4}$, we have performed $ab~initio$ 
calculations based on density functional theory (DFT) using the full-potential linearized-augmented
plane-wave (FP-LAPW) method as implemented in the WIEN2k code\cite{WIEN2k}.
Our calculations are based on generalized gradient approximation using the exchange-correlation functional proposed by Perdew, Burke and Ernserhof (PBE-GGA)\cite{pbegga}.
We used the value of $R_{\rm{MT}}K_{\rm{max}}=7$ in this study , where $R_{\rm{MT}}$ means the muffin tin sphere radius and $K_{\rm{max}}$ means the largest wave vector. The muffin tin sphere was chosen to be $R_{\rm{MT}}^{\rm{(Mn)}}=2.00 $ a.u., $R_{\rm{MT}}^{\rm{(Mg)}}=1.50$ a.u., and $R_{\rm{MT}}^{\rm{(O)}}=1.20$ a.u..
For SCF calculations, we used the $\bm{k}$-mesh of size $10\times10\times10$ $k$-points obtaind by Monkhorest-Pack sampling method\cite{monkhorst} and energy convergence criteria were set as $10^{-6}{\rm{Ry}}$.

To evaluate the transverse transport property of MnMg$_{2}$O$_{4}$, we calculated AHC ($\sigma_{xy}$) from following Kubo formula
\begin{align}
    \sigma_{xy} &= \frac{1}{(2\pi)^3}\sum_n \int_{\rm{BZ}} d\bm{k} f(E_{n\bm{k}}) \Omega_n^{z}(\bm{k}), \\
    \Omega_n^{z}(\bm{k}) &=-\frac{i e^2}{\hbar}\sum_{m(\neq n)}
\frac{\bra{u_{n,\bm{k}}} v_x \ket{u_{m,\bm{k}}}\bra{u_{m,\bm{k}}} v_y \ket{u_{n,\bm{k}}}-\rm{c.c.}}{(E_n(\bm{k})-E_m(\bm{k}))^2},
\end{align}
where $f(E_{n,\bm{k}})$ is Fermi-Dirac distribution function, ${\Omega}^z$ is z component of Berry curvature, and $v_i(i=x,y)$ are velocity operators. To perform BZ integration, we adopted finer  $\bm{k}$-mesh of size $60\times60\times60$ $k$-points.
For a deeper understanding of topological nature of MnMg$_{2}$O$_{4}$, we analyzed effective model obtaind by maximally localized generalized Wannier functions (MLWF)\cite{wannierRMP} based on the results of DFT calculations. MLWFs were constructed to reproduce DFT results by using WIEN2WANNIER package\cite{WIEN2wannier} and WANNIER90 package\cite{wannier90}. Based on MLWF, we analyzed distribution of nodes and Berry curvature in $\bm{k}$-space by using WANNIERTOOLS package\cite{wanniertools}. 

We studied bulk MnMg$_{2}$O$_{4}$ with inverse-spinel structure, whose conventinal crystal structure is shown in Fig. \ref{struct}, which was produced with {\footnotesize VESTA}\cite{VESTA}. The crystal structure belongs to a face-centered cubic Bravais lattice with space group $\rm{F\overline{d}3m}$ (No.227). As shown in Fig. \ref{struct}, Mn atom is surrounded by four O atoms , which acts as tetrahedral field. 
We performed optimization to obtain lattice constant and internal atomic positions.
The optimized lattice constant is $a=b=c=8.380$ \AA. Among 56 atoms MnMg$_{2}$O$_{4}$ in a unit cell,  Mn atoms locate at $8a(0.125,0.125,0.125)$ Wyckoff positions, Mg atoms locate at $16d(0.500,0.500,0.500)$ Wyckoff positions, and O atoms locate at $32e(0.251,0.251,0.251)$ Wyckoff positions, where we chose origin choice 2. These values are used in DFT calculations.



\section{Results}

\begin{figure}
\includegraphics[clip,width=9cm]{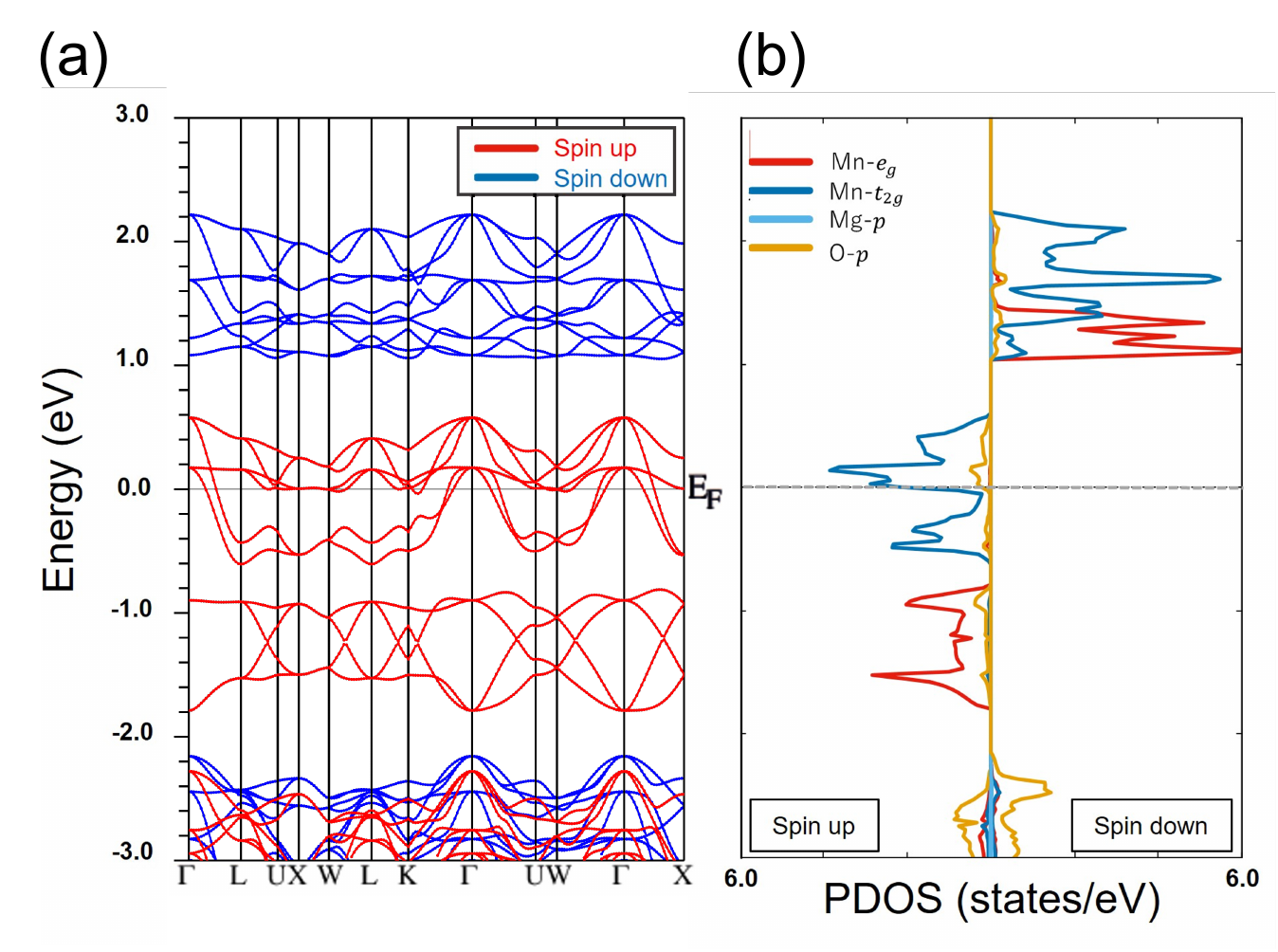}
\caption{
Calculated (a) electronic band structures and (b) partial density of states for Mn $e_{g}$ and $t_{2g}$, Mg $2p$, and O $2p$ orbitals in MnMg$_{2}$O$_{4}$.
}
\label{bandDOS}
\end{figure}
First, we show the band structure of MnMg$_{2}$O$_{4}$ along the main symmetry directions
in Fig. \ref{bandDOS} (a).
It can be clearly seen that the energy bands near the Fermi level are all in the spin-up channel.
We also show the orbital projected density of states (DOS) in majority-spin (positive) and minority-spin
(negative) channels in Fig. \ref{bandDOS} (b).
Now the semimetallic feature with an about 3eV minority spin gap of MnMg$_{2}$O$_{4}$ is clearly
noticeable.
One can see 
that the energy band crossing the Fermi level mainly consists 
of $t_{2g}$ orbitals of Mn 3d-electrons.
Also, 
the fully spin-polarized $e_{g}$-band with the hourglass dispersion is found about 1eV below
the Fermi level.
The results shown in Fig. \ref{bandDOS} confirm our expectation that the high-spin state is realized
in MnMg$_{2}$O$_{4}$.

In Fig. \ref{bandDOS}, We also notice the doubly degenerate bands with a very flat 
dispersion 
along the $X-W$ symmetry line just above the Fermi energy.
Since it is known that topological line nodes appear along the $X-W$ path in the single-orbital
diamond lattice, we can expect that these degenerate bands can be a manifestation of such 
nodal lines in this system.
\begin{figure}
\includegraphics[clip,width=7cm]{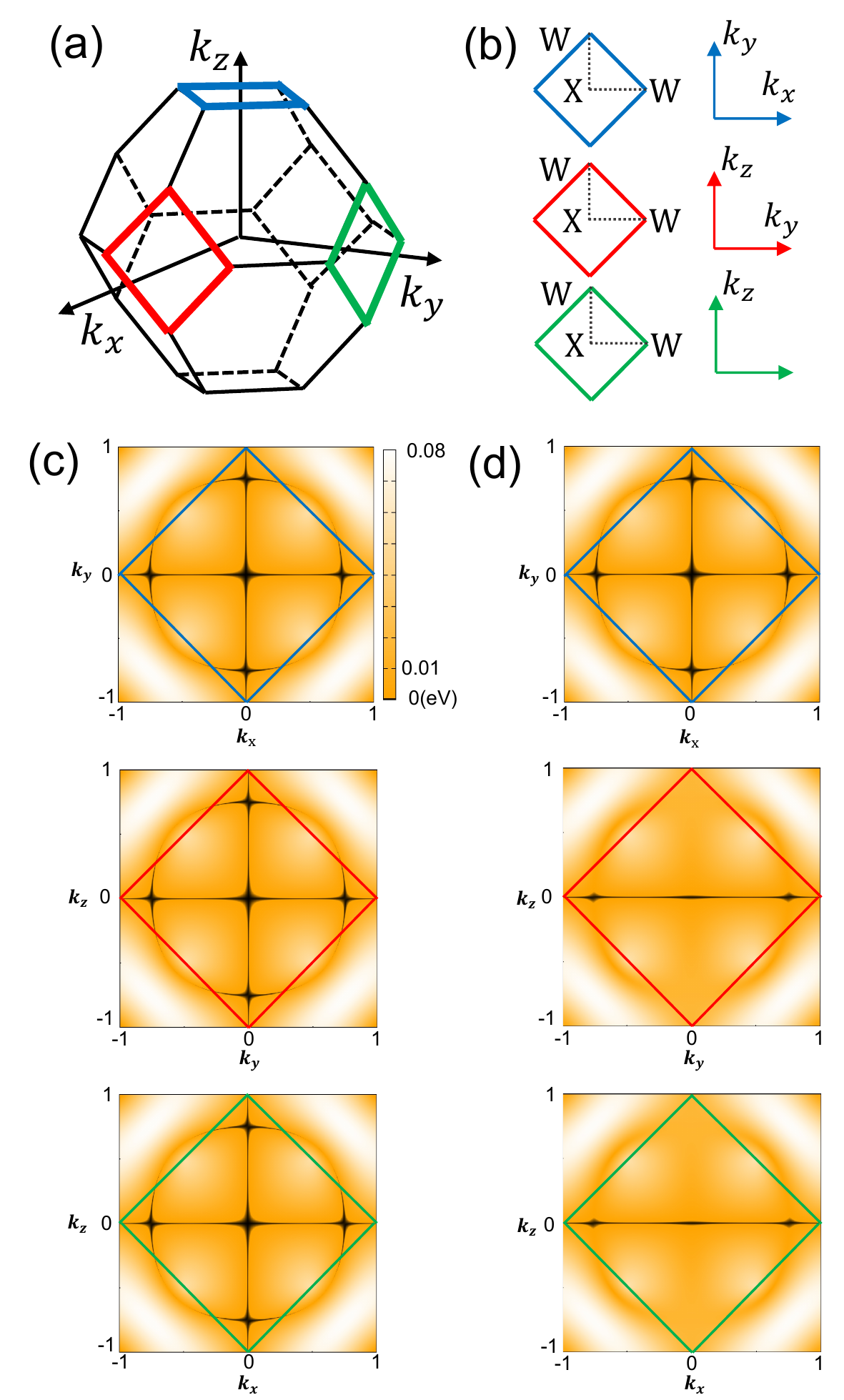}
\caption{
(a) First Brillouin zone (BZ) of the inverse spinel structure shown in Fig. \ref{struct}.
The three square $k$-planes bordered by blue, red, green lines
are the (001), (100), and (010) surfaces at the Brillouin zone
edge, respectively. 
The $X-W$ symmetry lines on these surfaces are depicted in (b). 
The energy gap between the two entangled bands on these planes
calculated without and with the spin-orbit interaction is shown
in (c) and (d), respectively.
The blue, red, green borders in (c) and (d) correspond to those in (b).
}
\label{nodalline}
\end{figure}
Thus, it is interesting to look at the energy gap between these two bands near the $X-W$ symmetry lines
in the Brillouin zone.
The first Brillouin zone of the system is shown in Fig. \ref{nodalline} (a);  
the three square $k$-planes bordered by blue, red, green lines are the (001), (100), 
and (010) surfaces at the Brillouin zone edge, respectively.
These $k$-planes contain the $X-W$ symmetry lines as shown in Fig. \ref{nodalline} (b).
We show the energy gap between the two entangled bands on these $k$-planes obtained without and with the spin-orbit interaction in Fig. \ref{nodalline} (c) and (d). 
In the absence of the spin-orbit interaction, these three $k$-planes are equivalent, and the system has two equivalent nodal lines crossing each other on each of the planes, as shown in Fig. \ref{nodalline} (c).
On the other hand, if we take into account the spin-orbit 
interaction and let the magnetization direction be (001), the three $k$-planes are no longer equivalent, and only one nodal line remains on the (100) and (010) planes, as shown in Fig. \ref{nodalline} (d). 
Thus we confirm that MnMg$_{2}$O$_{4}$ is approximately a nodal line semimetal.

\begin{figure}
\includegraphics[clip,width=8cm]{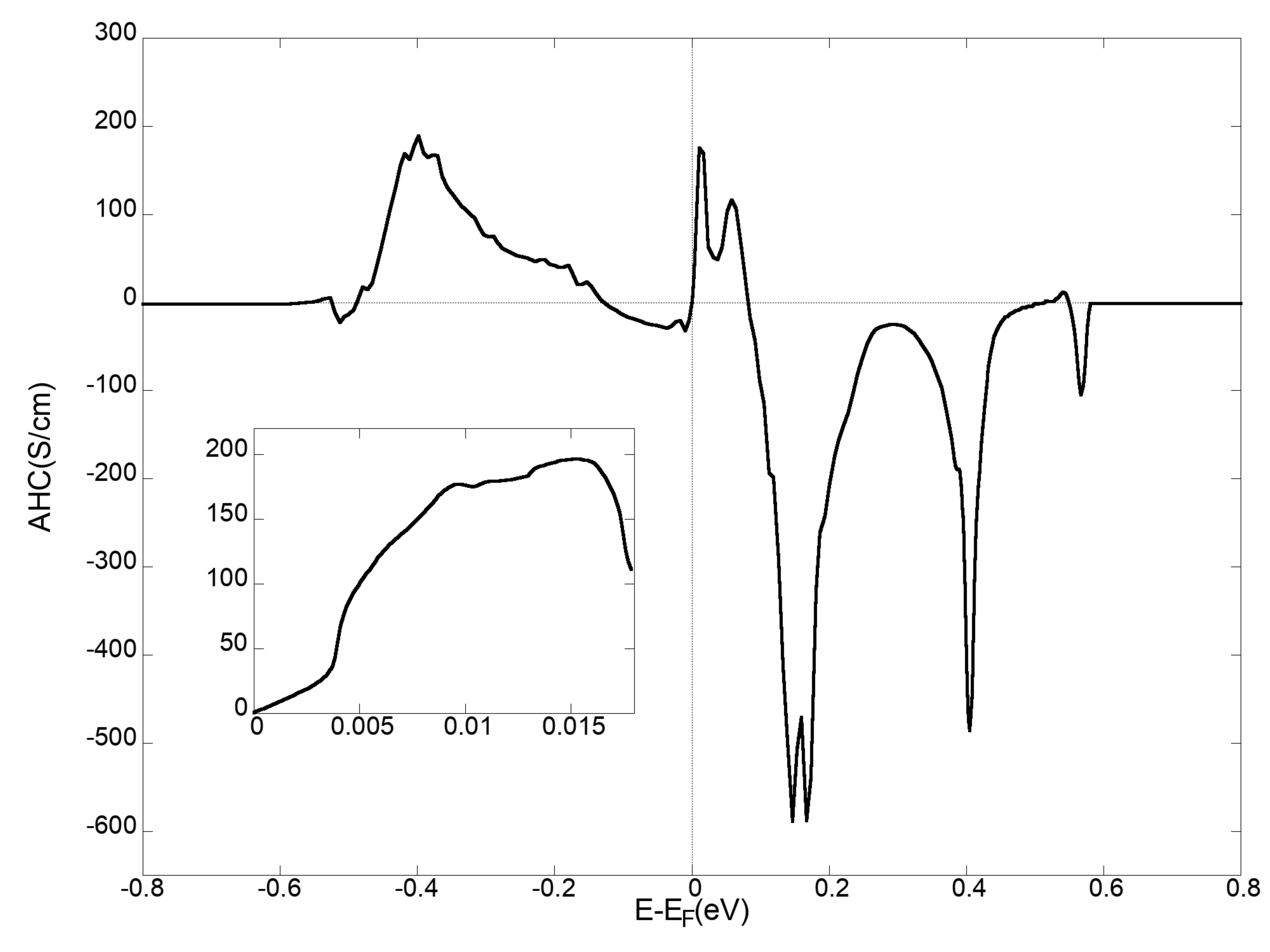}
\caption{
Chemical potential dependence of calculated AHC. 
The inset shows the detailed plot of AHC in the range of
$E=0$-$0.018$eV.
}
\label{ahc}
\end{figure}
Next, let us look at the intrinsic AHC $\sigma_{xy}$ as a function of the chemical potential of the system when the magnetization is along the [001] axis, assuming a rigid band structure.
We compute the intrinsic AHC using the Kubo formula in the clean limit.
We use a finer $k$-mesh with 14671 $k$-points 
within the reduced Brillouin zone for the AHC calculation using the OPTIC module in WIEN2k code.
In Fig. \ref{ahc}, we show the result of the AHC as a function of the chemical potential.
Although the intrinsic AHC is only 0.8 S/cm on the Fermi level, it increases sharply with slightly increasing the chemical potential from $E_{\mathrm F}$, then the AHC shows a peak value of $\sigma = 200$ S/cm at $E-E_{\mathrm F} = 0.015$ eV, as seen in the inset of Fig. \ref{ahc}. 

It is tempting to think here that the large AHC at around $E-E_{\mathrm F} = 0.015$ eV is 
due to the topological nature of the nodal lines found in Fig. \ref{bandDOS}. 
Thus finally, we analyze the Berry curvature distribution on the $k$-planes 
defined in Fig. \ref{nodalline} to confirm the effects of the nodal lines on the AHC.
%
\begin{figure}[h]
\includegraphics[clip,width=9cm]{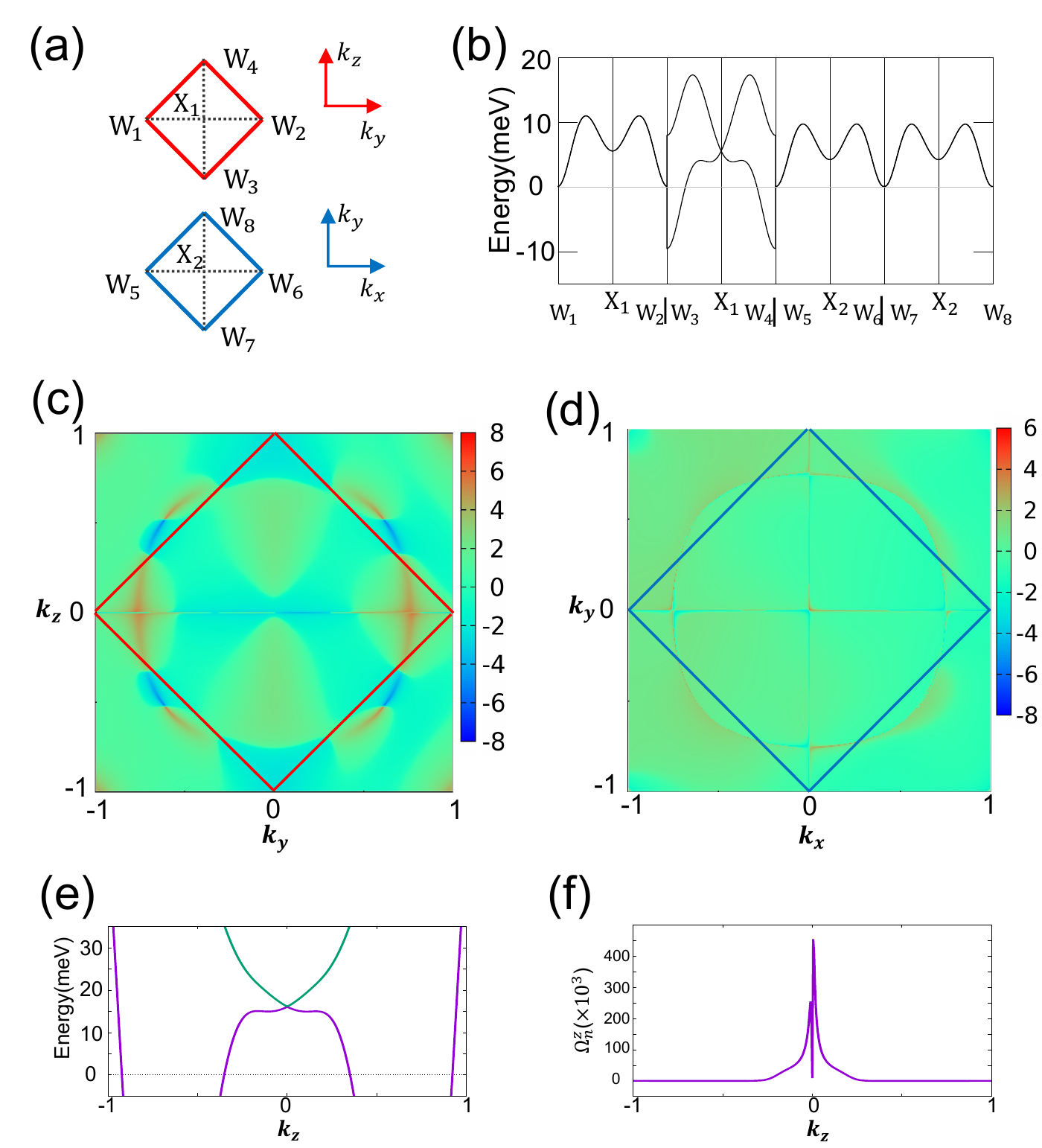}
\caption{\label{fig:epsart} 
(a) The (100) and (001) $k$-planes as in the same manner in Fig. \ref{nodalline} (b). 
(b) The energy dispersion on $X-W$ symmetry lines, where the labels of symmetry points are given in (a). 
The Berry curvature on the (100) and (001) $k$-planes are shown
in (c) and (d), respectively.
In (c) and (d), the magnitude of the $z$ component of the Berry curvature is plotted in a logarithmic scale  $\rm{sign}(\Omega_n^{z})*(1+\rm{{log}_{10}}(\Omega_n^{z}))$}.
(e) The band dispersion along $k_{y} = 0.75$ near the (100) $k$-plane.
(f) Calculated Berry curvature along $k_{y}=0.75$ on the (100)
$k$-plane.
\label{berry}
\end{figure}
%
Figure \ref{berry} (a) shows the (100) and (001) $k$-planes
as in the same manner
in Fig. \ref{nodalline} (a) and (b).
As mentioned earlier, if the magnetization of the system is along the [001] direction, 
the electronic states of these two planes are not equivalent due to the effect of spin-orbit interaction.
Thus for clarity, we here distinguish the $X$ and $W$ 
points which were originally equivalent with subscripts
as in Fig. \ref{berry} (a).
A detailed view of the band dispersion along the symmetry lines on these planes is shown
in Fig. \ref{berry} (b).
The narrow energy band is found to be in the range of -10 to 18meV, which coincides with 
the peak value of the AHC.
Furthermore, we find the hot spots of the Berry curvature around $k_{y}\simeq \pm 0.75$ 
on the nodal line on the (100) $k$-plane as shown in Fig. \ref{berry} (c).
Note here that no hot spots or prominent features are found in the calculated
Berry curvature on the (001) $k$-plane as can be seen from Fig. \ref{berry} (d).
Let us take a closer look at the band structure and the Berry curvature distribution 
around the hot spots.
Figure \ref{berry} (e) and (f) shows the band dispersion along $k_{y}=0.75$ near the (100) $k$-plane.
A nearly flat portion of the band with a band crossing at $k_{z}=0$ lies around $E=0.015$eV, which coincides again with the peak value of the AHC 
in Fig. \ref{ahc}.
The calculated Berry curvature along $k_{y}=0.75$ on the (100) $k$-plane
is also shown in Fig. \ref{berry} (f).
We notice that the Berry curvature is large in the flat portion of the band
and exhibits strong peaks close to the crossing point.
It should be noted here that the Berry curvature is almost zero at $k_{z}=0$,
i.e. on the nodal line.
Therefore, we may conclude that the large AHC found at $E\sim 0.015$eV
is not attributed to the nodal lines themselves, but to the hot spots
found in very close proximity to them.
More detailed studies on the origin of the hot spots and their topological nature will be reported in a forthcoming paper.

\section{Summary}
In summary, we have investigated the electronic states and the AHC
in the inverse spinel compound MnMg$_{2}$O$_{4}$ based on first-principles calculations, 
stimulated by the recent theoretical works of VMg$_{2}$O$_{4}$.
The results have confirmed that this material is 
a half-metallic semimetal with the line nodes
slightly above the Fermi level, and also exhibits fairly large AHC.
The Berry curvature distribution near the nodal
lines in the Brillouin zone has been also studied, and no direct relationship between the large AHC and the nodal lines was found.
Instead, the hot spots of Berry curvature are found in very close proximity to the nodal lines.
Future work is required to reveal the  topological nature of the hot spots and
the origin of the large AHC.



\begin{acknowledgments}
This work was supported by JSPS 
KAKENHI Grant Number JP21H01025 and JP20H01830,
and also by JST CREST Grant Number JPMJCR17J5 and JPMJCR18T2.
Some numerical computations were carried out at the Cyberscience Center, Tohoku University, Japan.
\end{acknowledgments}

\section*{Data Availability}
The data that support the findings of this study are available from the corresponding author upon reasonable request.

\nocite{*}
\providecommand{\noopsort}[1]{}\providecommand{\singleletter}[1]{#1}%
%

\end{document}